\documentclass[%
 aip,
 sd,%
 amsmath,amssymb,
 reprint,%
]{revtex4-1}

\usepackage{graphicx}
\usepackage{dcolumn}
\usepackage{bm}
\usepackage{multirow}
\usepackage{relsize}
\usepackage{array}
\usepackage{units}
\usepackage{ctable}
\usepackage{verbatim}
\usepackage{times}
\usepackage{amsmath}
\usepackage{amsfonts}
\usepackage{amssymb}
\usepackage{verbatim}
\usepackage{amsfonts}

\usepackage{pifont}

\begin{document}

\newcommand{\Heff}{\mathcal{H}_\mathrm{eff}}
\newcommand{\vr}{{\bf x}}
\newcommand{\vR}{{\bf r}}
\newcommand{\grad}{\nabla}

\preprint{AIP/123-QED}

\title{Intra-chain organisation of hydrophobic residues controls inter-chain 
aggregation rates of amphiphilic polymers}

\author{Patrick Varilly}
\affiliation{Department of Chemistry, University of Cambridge, Lensfield Road, 
Cambridge CB2 1EW, UK}
\author{Adam~P.~Willard}
\affiliation{Department of Chemistry, MIT, Cambridge MA, USA}
\author{Julius~B.~Kirkegaard}
\affiliation{Department Applied Mathematics and Theoretical Physics, University 
of Cambridge, Centre for Mathematical Sciences, Wilberforce Road, Cambridge CB3 
0WA, UK}
\author{Tuomas~P.~J.~Knowles}
\affiliation{Department of Chemistry, University of Cambridge, Lensfield Road, 
Cambridge CB2 1EW, UK}
\affiliation{Cavendish Laboratory, Department of Physics, University of 
Cambridge, J J Thomson Avenue, Cambridge CB3 0HF, UK}
\author{David Chandler}
\affiliation{Department of Chemistry, University of California, Berkeley, 
California 94720, USA}

\begin{abstract}
Aggregation of amphiphiles through the action of hydrophobic interactions is a 
	common feature in soft condensed matter systems and is of particular 
	importance in the context of biophysics as it underlies both the generation 
	of functional biological machinery as well as the formation of pathological 
	misassembled states of proteins. Here we explore the aggregation behaviour 
	of amphiphilic polymers using lattice Monte-Carlo calculations and show that 
	the distribution of hydrophobic residues within the polymer sequence 
	determines the facility with which dry/wet interfaces can be created and 
	that such interfaces drive the aggregation process.
\end{abstract}



\maketitle

Due to their importance in governing self-assembly of biological components, 
hydrophobic interactions and the mechanism of hydrophobic collapse leading to 
the aggregation of hydrophobic species in an aqueous environment have been 
studied in detail using approaches ranging from spectroscopy to atomistic and 
coarse-grained simulations~\cite{wallqvist1995computer, 
lum1999hydrophobicity,huang2001scaling,raschke2001quantification, 
huang2002hydrophobic,dixit2002molecular,chandler2005interfaces, 
granick2008chemistry,rasaiah2008water,WillardChandler2008, 
berne2009dewetting,patel2010fluctuations,hummer2010molecular, 
garde2011unraveling,patel2012sitting,ben2015hydrophobic}.  The phenomenon of 
hydrophobic collapse by its very nature involves the removal of water molecules 
between adjacent hydrophobic entities in order to allow for them to come 
together, and therefore the creation of an interface with unsatisfied 
hydrogen-bonding separating "wet" solvent from "dry" aggregated hydrophobes in a 
manner reminiscent of a liquid-vapour phase transition. The picture that has 
emerged from computational studies of the collapse of hydrophobic chains is that 
it is the creation of such interfaces which controls the transition between 
solvated hydrophobes and their compact aggregated 
state~\cite{tenWoldeChandler2002,miller2007solvent,jamadagni2008interfaces}.  In 
biological systems, hydrophobic units rarely occur in isolation and are most 
commonly part of a macromolecular system. In the present work, we investigate 
using lattice Monte-Carlo calculations the nature of the hydrophobic collapse 
for polymers with a varying distribution of hydrophobic and hydrophilic elements 
and demonstrate that the clustering of hydrophobic entities is crucial for 
nucleating the formation of dry interfaces driving the eventual aggregation 
process.

\section{Off-lattice solutes in a lattice gas solvent model}

\label{sec:LLCW:MovingSolutes}
Hydrophobic assembly is characterised by the expulsion of water molecules from 
aggregates of hydrophobic entities. This effect can be captured by considering 
the evolution of the local water density field. The short length scale density 
fluctuations contributing to the local density are characterised by rapid 
relaxation and follow to a very good approximation Gaussian 
statistics~\cite{hummer1996information}. These short scale fluctuations can 
therefore be integrated analytically~\cite{chandler1993gaussian, 
tenWoldeChandler2002,varilly2011improved}, resulting in a coarse grained density 
field $\rho(\vec{r})$ which is readily simulated with a discretized binary field 
$n_i$ which tracks the density fluctuations resulting in the appearance of cells 
$i$ with a lower density, "vapour" cells $n_i=0$, and "wet" cells within the 
bulk solvent with $n_i = 1$.  This description is particularly well suited for 
numerical evaluation as the computationally costly short length scale 
fluctuations characteristic of atomistic models have been treated analytically.  
Within this picture, there is a cost to create a wet/dry interface, given by 
nearest neighbour interactions of the form $n_i n_{i\pm 1}$ and an energy 
associated with the solvation of chemical species $n_i \mu_{ex}$.  This 
description is therefore equivalent to a 3D lattice gas system with the 
Hamiltonian:
\begin{equation}
\mathcal{H}[\{n_i\}, \{h_i\}] = \sum_i [-\mu + \mu_\mathrm{ex} h_i]n_i + \epsilon \sum_{\langle i,j\rangle} n_i n_j
\end{equation}
where $\mu$ is the solvent chemical potential and $\langle \dots \rangle$ 
indicates summation over nearest neighbours on the lattice.  The presence of 
hydrophobic solutes at lattice sites $i$ with $h_i = 1$ results in an excess 
chemical potential $\mu_\mathrm{ex}$. 

The values of the parameters governing the coarse grained water degrees of 
freedom, $\mu = 3\epsilon - 2.25\cdot 10^4\, k_B T$ and $\epsilon = 1.51\,k_B 
T$, can be determined for a lattice size of $l=0.21$ nm through comparisons with 
the experimental bulk values for the isothermal compressibility and the surface 
tension $\gamma = \epsilon/(2l^2) = 0.07\mathrm{N\,m^{-1}}$ of water at room 
temperature and 1 atm pressure. The value of $\mu_\mathrm{coex} = 3 \epsilon$ 
represents the chemical potential of the solvent at phase coexistence with the 
vapour phase, and the small difference $\mu-\mu_\mathrm{coex} \ll k_B T$ 
highlights the fact that water is close to phase coexistence under standard 
conditions. It has been shown that this description of water reproduces 
faithfully the key properties associated with hydrophobic interactions, in 
particular the characteristic solvation free energy changes with increasing 
solute sizes. This coarse-grained water description has previously been used to 
study the collapse of a single hydrophobic chain \cite{tenWoldeChandler2002}, 
and we extend this approach here to cover the aggregation of amphiphilic chains. 

With the specific parameterization given above, the lattice solvent used here is below the roughening transition for the three-dimensional Ising model.
As a result, there can be lattice artifacts due to a tendency of an
interface to align with the orientation of the underlying lattice vectors 
\cite{Weeks1977}.  A true liquid-vapor interface would not exhibit this 
behavior.  Vaikuntanathan and Geissler have recently demonstrated that this 
tendency can give rise to inaccurate solvation free energies for hydrophobic 
solutes that are nanometer sized or larger \cite{Vaikuntanathan2014}.  This 
inaccuracy is most significant for aspherical or irregularly shaped solutes, but 
grows less pronounced as the roughening transition is approached from below. In 
the case presented here the model is only slightly below the roughening 
transition, which is located at $\epsilon\approx 1.64 k_\mathrm{B} T$, and so 
lattice artifacts are only expected to manifest on length scales of about $1-2 
\ell$. Since the critical nucleus size for hydrophobic peptide aggregation is 
about 1nm, or $\approx 5\ell$, we expect any lattice artifacts associated with 
being below the roughening transition to be negligible for the results presented 
here.
Indeed, the mechanism for collapse found for a hydrophobic chain with the 
lattice solvent model we employ\cite{tenWoldeChandler2002} is consistent with 
that found for same hydrophobic chain with an atomistic solvent 
model.\cite{miller2007solvent} For more generally shaped solutes, effects of 
lattice artifacts may best be avoided by adopting Vaikuntanathan and Geissler’s 
related lattice model, \cite{Vaikuntanathan2014, Vaikuntanathan2016} which is 
slightly more complicated than that of Ref.~(\onlinecite{tenWoldeChandler2002}).

For the hydrophobic segments, the excess chemical potential is given by 
$\mu_\mathrm{ex} = c_{\text{phob}} v$, where $c_{\mathrm{phob}} = 60 \,k_B T 
\,\mathrm{nm^{-3}}$ is taken to be the reversible work required to accommodate a 
hydrophobe of volume $v$. Idealised hydrophilic segments are water like and as 
such the excess chemical potential due to the presence of hydrophobic solutes 
vanishes $\mu_\mathrm{ex} = 0$. Furthermore, weak depletion forces act between 
two hydrophobic particles and originate from the reduction of volume from which 
the solute excludes solvent molecules.

\begin{figure}[tbh]
\includegraphics[width=0.5\textwidth]{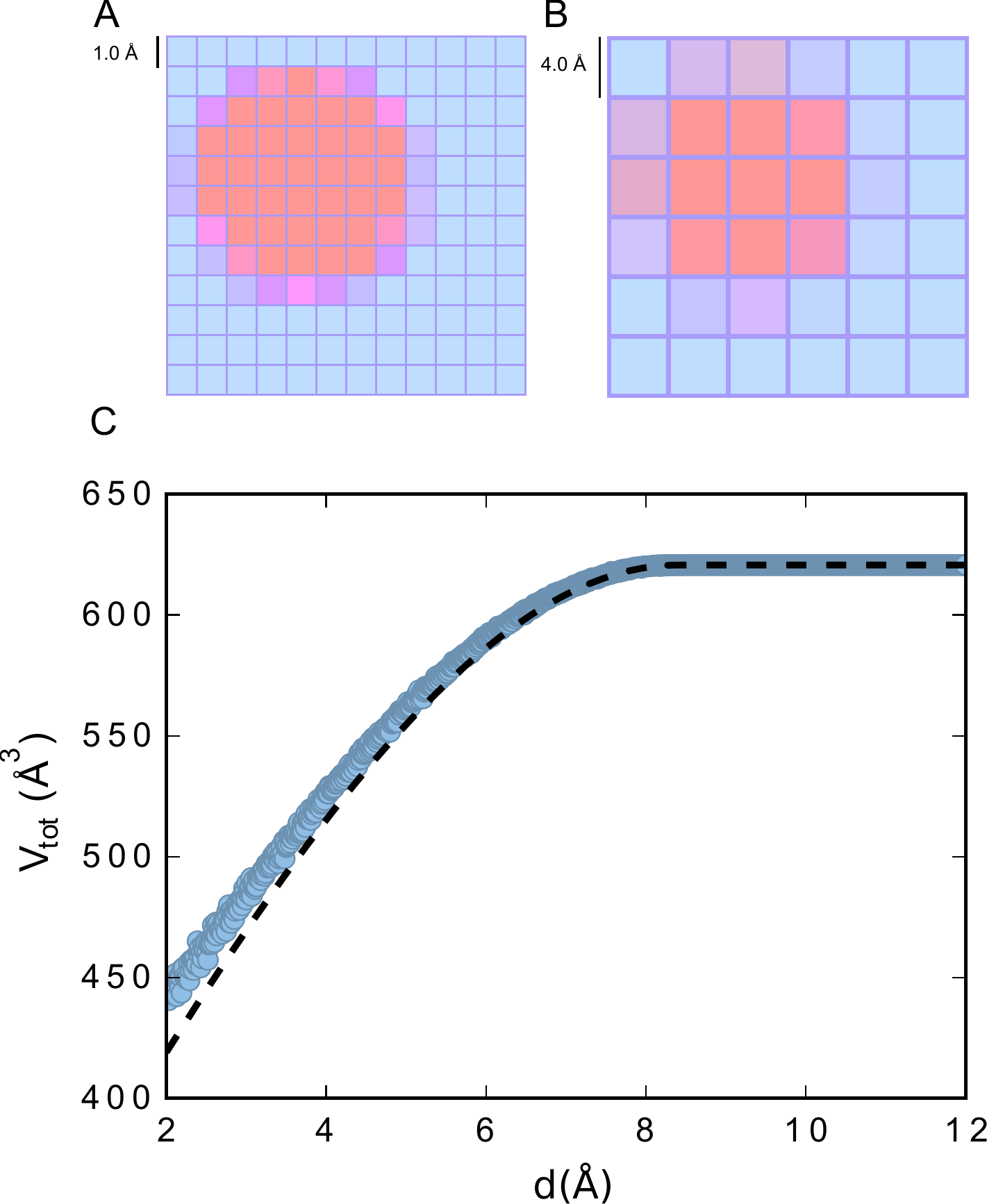}
\caption{
\label{fig:LLCW:movingR3p37}Shown in A, the fine grid on which the overlap 
volume is calculated. This is mapped onto the lattice on which the field $n_i$ 
is defined as shown in B. In C, the total volume excluded by two spheres of 
radius $R=4.20\,$\AA\ when a distance~$d$ apart, as calculated exactly 
(Equation~\eqref{eqn:LLCW:exactTwoSphere}) and by the numerical approximation 
scheme with spacing 1\AA.}
\end{figure}

The solvent degrees of freedom in the model, $\{n_i\}$, can be efficiently 
sampled using a Metropolis Monte Carlo algorithm. By contrast, sampling the 
solute degrees of freedom is more complex. The principal problem is to calculate 
the overlap volume~$v_a$ between the excluded volume~$v$ and a given fine 
cell~$a$.  The volume~$v$ is typically a union of overlapping spheres, one for 
each excluded volume associated with a solute particle, here a polymer segment.  
Previously in
Refs.~(\onlinecite{tenWoldeChandler2002})~and~(\onlinecite{WillardChandler2008}), 
an interpolation scheme was used that only works if no point in space is within 
the solvent-excluding radius of more than two spheres simultaneously, and a 
solute geometry was chosen that avoids this situation. In the present paper we 
focus on aggregation of multiple chains where densely packed structures are 
expected and thus multiple overlaps will occur -- thus the existing 
interpolation scheme is not suitable.

\begin{figure}[tbh]
\includegraphics[width=0.5\textwidth]{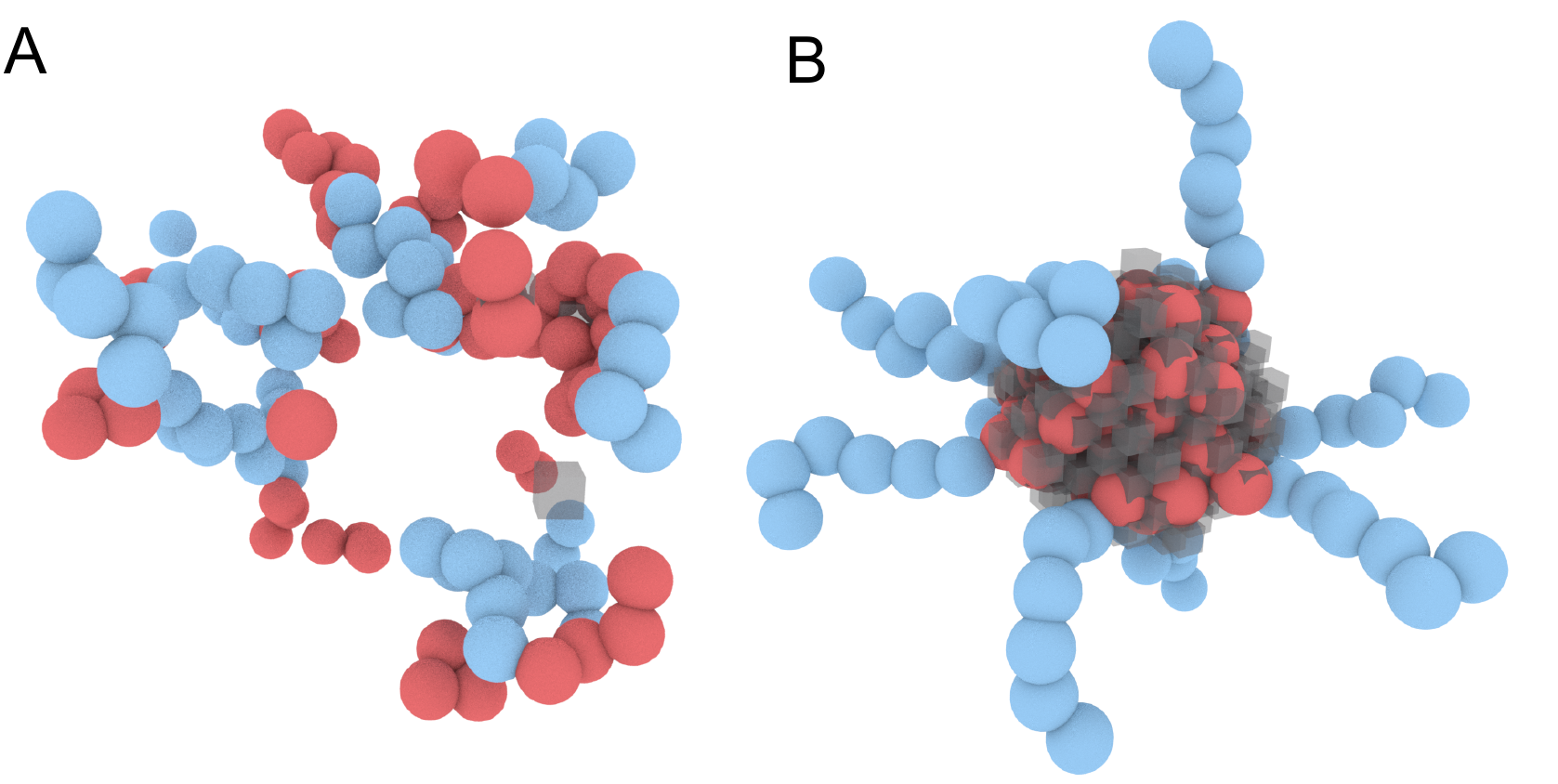}
\caption{
\label{fig:collapse}
Hydrophobic collapse of an amphiphilic chains. In A is shown the system before collapse and in B the state of the system after 20,000 MC moves have been performed. Vapour lattice sites are shown as grey, transparent cubes, hydrophobic residues in red, and hydrophilic residues in blue.}
\end{figure}

Here, we discuss a partial solution to the above problem. Specifically, we 
present an approximate method of calculating~$v_a$ when $v$ is, as above, a 
union of possibly overlapping spheres of a few different sizes. The gradient of 
$v_a$ with respect to the centers of these spheres is also easy to calculate.  
In principle, propagating these gradients to obtain gradients 
of~$\Heff[\{n_i\}]$ is then simply a (non-trivial) bookkeeping exercise.
In describing our scheme, we treat cell indices~$a$ as vectors that can be added 
and subtracted. We denote by $\vr_a$ the coordinates of the corner of cell~$a$ 
with lowest Cartesian components.  For a solvent-excluding sphere of radius~$R$ 
centered at~$\vr_0$, we can pre-calculate the overlap volumes~$\hat v_s$ of all 
cells~$s$ by any method, such as Monte Carlo integration.  We do this once at 
the beginning of a simulation.

Generically, the center~$\vR$ of a solvent-excluding sphere will not coincide with a cell corner. We denote the indices of the eight corners of the cell containing~$\vR$ by $a_1,\ldots,a_8$, and their positions by~$\vr_1,\ldots,\vr_8$.  We construct eight non-negative weights~$c_1(\vR),\ldots,c_8(\vR)$, the sum of which is~$1$ and whose value depends continuously on~$\vR$, such that
$\vR = \sum_{k=1}^8 c_k(\vR) \vr_k$.  Any scheme with these characteristics, such as trilinear interpolation, can be used. The overlap volumes for cells~$a$ near~$\vr$ are then estimated by
\begin{equation}
{\tilde v}_a(\vR) \approx \sum_{k=1}^8 c_k(\vR) \hat v_{a - a_k}.
\label{eqn:LLCW:tildeva}
\end{equation}
This interpolation scheme has the desirable property that the total volume of a 
sphere, given by $\sum_a {\tilde v}_a(\vR)$, is independent of~$\vR$.

\begin{figure}[tbh]
\includegraphics[width=0.5\textwidth]{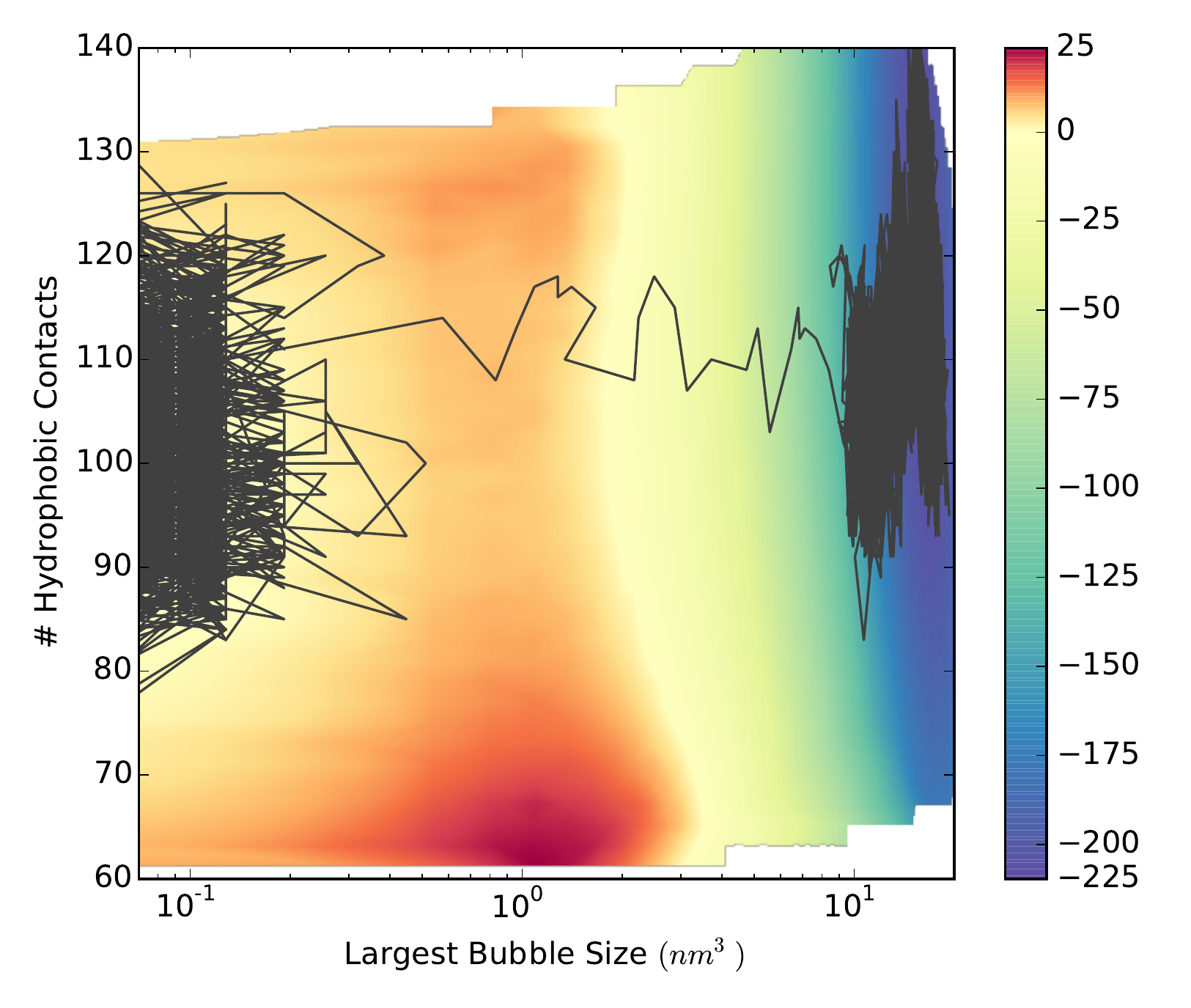}
\caption{
\label{fig:traj}Map of the free energy landscape for the aggregation of a 
solution of diblock polymer chains composed of six hydrophilic residues followed 
by six hydrophilic residues computed using umbrella sampling (colour scale: 
energy in $k_B T$).  Superimposed in black is shown an unbiased trajectory 
displaying aggregation.}
\end{figure}

When a solute is composed of multiple spheres, centered at~$\vR^N$, we simply add together the overlap volumes given by Equation~\eqref{eqn:LLCW:tildeva} for each solute, but we cap the sum at the total volume of each fine cell, $\lambda_f^3$.  In summary, we have
\begin{equation}v_a \approx \min\left[\lambda_f^3, \sum_{n=1}^N {\tilde
    v}_a(\vR_n)\right]
 \end{equation}
This scheme is exact whenever one or more spheres overlaps cell~$a$ completely, 
as well as when two spheres both overlap cell~$a$ but not each other.  When two 
or more spheres both partially overlap cell~$a$ and each other, our scheme 
mildly overestimates the overlap volume.
We have evaluated the precision of our scheme by calculating the total volume $V_{\text{tot}}$ of two spheres of radius~$R$ whose position is varied, and the two spheres are placed at arbitrary positions with respect to the fine grid.  The total volume can be calculated analytically when the two sphere centers are a distance~$d$ apart,
\begin{equation}V_{\text{tot}}(d) = \begin{cases}2\times\frac43\pi R^3,&d > 2R,\\2\times\frac43\pi R^3 \\ \quad - \frac{\pi}{12} (4R + d) ( 2R - d )^2,&\text{otherwise}.
\end{cases}
\label{eqn:LLCW:exactTwoSphere}
\end{equation}
The results of this comparison are shown in Figure~\ref{fig:LLCW:movingR3p37}C when $R = 4.20\,$\AA\ and the fine grid resolution is~$1\,$\AA.  As expected, the exact and numerical results agree closely.  Of equal importance, the spread in the numerical estimate of the total volume is small, which suggests that the lattice artifacts of our overlap-volume scheme are quite modest while offering a robust and computationally advantageous solution to the problem of computing overlap integrals for off-lattice solutes.

We note that if the solute needed to be propagated through some variant of 
molecular dynamics, such as Langevin dynamics~\cite{tenWoldeChandler2002}, the 
gradient of~$\Heff [\{n_i\}]$ with respect to solute positions would also be 
needed, and this is easy to calculate to the present approximation scheme.

\section{Results}

The framework presented in this paper allows the study of solutes that can move 
freely in space in combination with an effective and computationally tractable 
explicit solvent model that exploits the statistical mechanics of lattice gas 
models. Using this approach, we have probed the aggregation behaviour of 
amphiphilic polymers with differing sequences of hydrophilic and hydrophobic 
residues. We model the dimensions of our chains on that of polypeptide chains; 
the excluded volume of the residues $V=310.3 \, \text{\AA}^3$ with a core volume 
$V=92.0 \, \text{\AA}^3$, is chosen to match that determined experimentally for 
amino acids \cite{levitt1976simplified}.  Furthermore, the excess chemical 
potentials of solvation possess values which cover the range measured for 
hydrophilic and hydrophobic amino acids\cite{Roseman1988}. The polymer is 
modeled as a Gaussian chain, and the simulation box with periodic boundary 
conditions has a volume of 216.0 $\mathrm{nm}^3$. We used a polymer mass 
concentration of 79.6 mg/ml, a value comparable to the total protein 
concentration in many organisms.

\begin{figure}[tbh]
\includegraphics[width=0.5\textwidth]{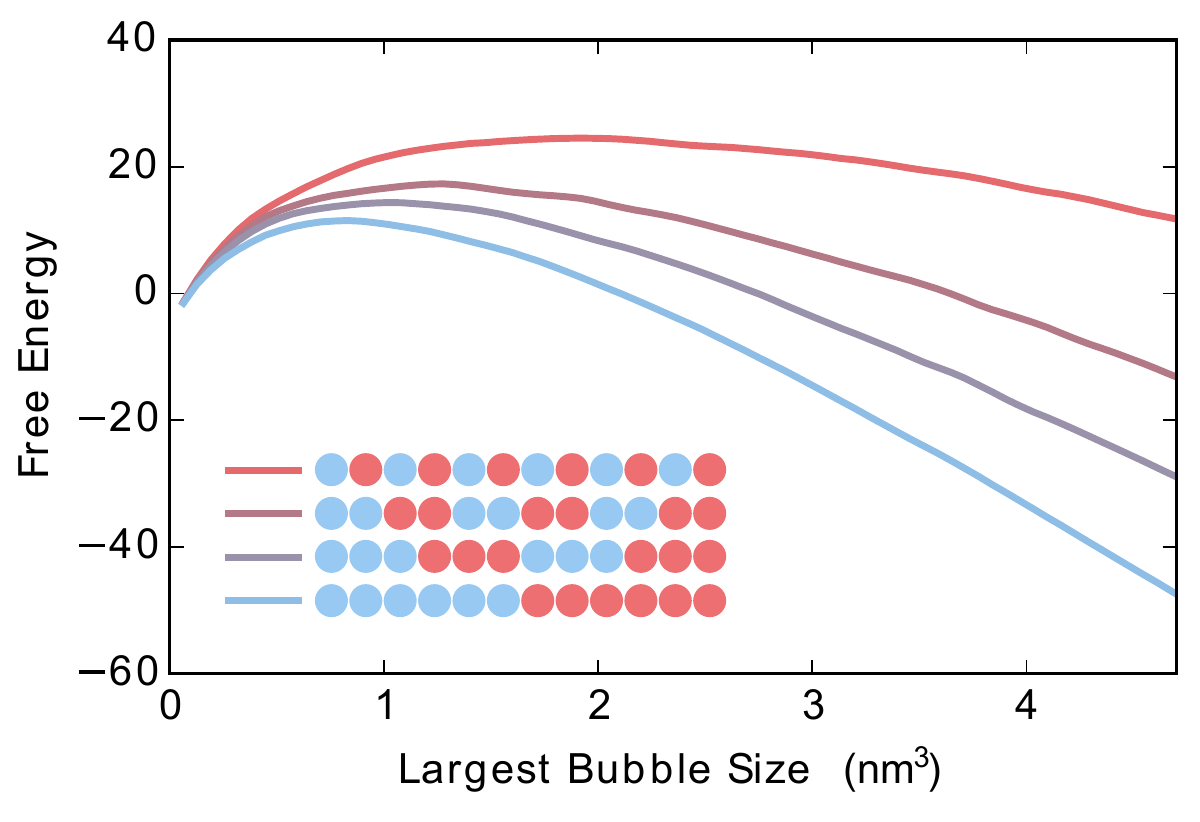}
\caption{
\label{fig:barrier}The free energy landscape in the solvent coordinate for polymers which all have as half of their residues hydrophobes and as the other half hydrophilic residues, but where the distribution of hydrophobes within the sequence has been varied.}
\end{figure}

\begin{figure}[tbh]
\includegraphics[width=0.5\textwidth]{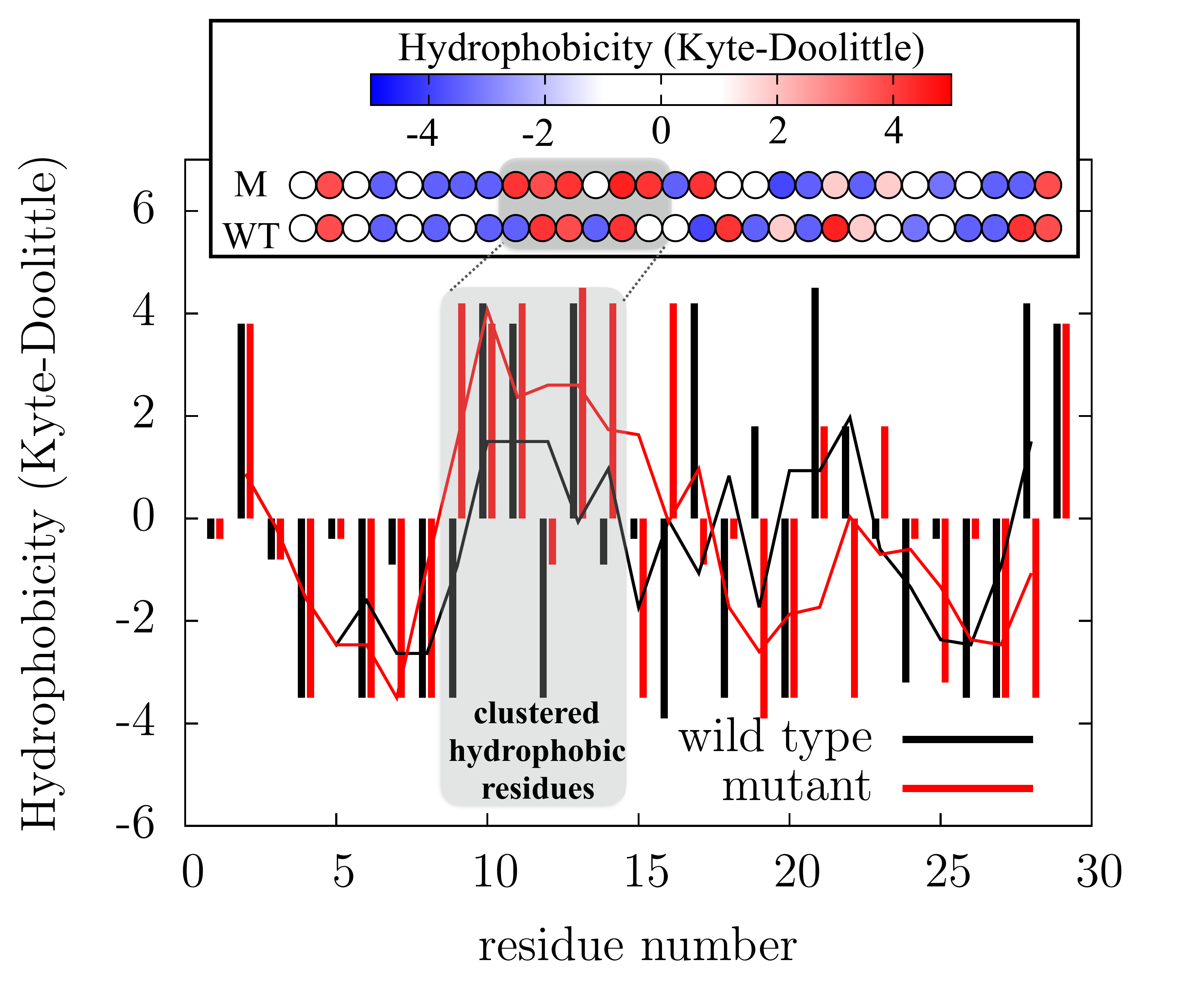}
\caption{
	\label{fig:exp}Analysis of the aggregation propensity horse heart 
	apomyoglobin and a scrambled sequence\cite{Monsellier2007a} (top).  The 
	local Kyte-Doolittle hydrophobicity is shown as a bar chart as a function of 
	residue number for both the wild type and scrambled sequence; a sliding 
	window average of size 4 over the sequence length of wild type horse heart 
	apomyoglobin (solid lines) compared to scrambled sequence resulting in an 
	aggregation prone mutant (red line), highlighting the role of clusters of 
	hydrophobic residues in favouring aggregation even in the absence of an 
	overall increase of the hydrophobicity of the sequence, in agreement with 
	the simulation results in Fig.\ref{fig:barrier}.  The sequences are from 
	\cite{Monsellier2007a}.}
\end{figure}

First, a solution of polymer chains composed of six hydrophobic residues 
followed by six hydrophilic residues spontaneously aggregates during unbiased 
simulations to form a cluster as shown in Fig.~\ref{fig:collapse} where the 
hydrophobic sections form a dry core and the hydrophilic segments are solvated 
on the outside of this core. We follow the mechanism of aggregation of the 
chains by focusing on two coordinates: a solvent coordinate measuring the size 
of largest bubble of vapour sites, and a polymer coordinate which describes the 
number of hydrophobic residues that are in contact, defined as their centres 
lying within a distance less than $2R + 1 \text{\AA}$, where $R$ is the residue 
radius. In this manner we have a reporter for both the changes in the 
conformations of the polymer chains as well as on the behaviour of the solvent.  
Unbiased trajectories show that during the aggregation process initially rapid 
fluctuations in the number of hydrophobic contacts are observed; this process in 
itself does not, however, lead to aggregation of the chains since any contacts 
formed are able to dissociate readily during the simulation. However, the system 
may undergo a critical fluctuation in the water coordinate leading to the drying 
of a hydrophobic contact. This fluctuation then drives the complete aggregation 
as other residues are subsequently recruited into this hydrophobic core. This 
mechanism is analogous to that observed for the formation of intra-molecular 
contacts in purely hydrophobic chains studied 
previously~\cite{tenWoldeChandler2002}.

The importance of fluctuations in the solvent degrees of freedom, which allow 
hydrophobic contacts to be stabilised, raises the question of how the ease of 
generating such fluctuations depends on the sequence in which hydrophobic and 
hydrophilic residues are distributed within the polymer chain. This question is 
also motivated by the empirical observation that there is evidence from studies 
of protein sequences that significant evolutionary pressures govern the 
distributions of hydrophilic and hydrophobic residues in order to avoid unwanted 
aggregation \cite{DuBay2004a}, in addition to the more conventional role of 
sequence in determining the final fold of the chain \cite{Anfinsen1973}. 

We investigated this question by generating different sequences of polymers 
which all share the same average composition of 50\% hydrophilic residues and 
50\% hydrophobic residues. The evaluation of the free energy barriers in 
Fig.~\ref{fig:barrier} against aggregation reveals that the sequence of the 
polymer, even for a constant average composition, has a major role in 
determining its propensity to aggregate. The free energy barriers are 
significantly larger for polymers which have only small hydrophobic clusters and 
where the residues are evenly distributed throughout the chain. By contrast, for 
polypeptide chains where the hydrophobic residues are segregated to one end of 
the chain, we observe a significantly reduced barrier and increased propensity 
to aggregate. The entropic penalty of bringing together a critical number of 
hydrophobic residues, that are required to observe a drying transition leading a 
stable hydrophobic contact, is significantly greater when these residues are not 
at adjacent positions in the chain but distributed throughout the sequence. In 
this manner, the interplay between polymer degrees of freedom and solvent 
degrees of freedom generates a very significant and sensitive dependency of the 
aggregation potential of the chain on the precise placement of the hydrophobic 
residues.

Interestingly, experiments designed to probe the aggregation propensity of 
sequence scrambled variants of the first 29 residues of horse heart apomyoglobin 
have been reported\cite{Monsellier2007a}. This system consists of a short 
peptide where the amino acid composition has been kept constant, but several 
mutants were generated where the position of the amino acids within the sequence 
was varied. It was observed that such mutants possessed markedly different 
aggregation propensities despite their common amino acid composition, with 
aggregation prone clusters of hydrophobic residues being particularly associated 
with a high propensity to aggregate. To facilitate a quantitative comparison 
between these different peptides we map their sequences onto a quantitative 
measure of hydrophobicity. The Kyte-Doolittle scale\cite{Kyte1982} is one such 
measure that associates a scalar hydrophobicity score with each amino acid.  If 
we consider the local Kyte-Doolittle hydrophobicity\cite{Kyte1982} of the most 
aggregation prone mutant reported in the study relative to the aggregation 
resistant wild type, as shown in Fig.~\ref{fig:exp}, a marked difference in the 
distribution pattern of the hydrophobic residues is apparent, with the 
aggregation-prone mutant possessing a cluster of hydrophobic residues which are 
distributed more evenly throughout the sequence for the wild type.  This type of 
observation is in close agreement with the importance of clusters of hydrophobic 
residues determined from simulations in the present work.

\section{Conclusions}
In this paper we have developed and described an approach to use off-lattice 
solutes within a statistical mechanical model of lattice solvents. Our system is 
computationally tractable, yet includes the relevant solvent degrees of freedom.  
We have used this system to probe the role of the distribution within the 
sequence of amphiphilic polymers of the positions of hydrophobes. Our 
simulations show that highly aggregation prone chains result when the 
hydrophobes are not distributed evenly within the chain but cluster in close 
proximity along the chain. This clustering favours the nucleation of a dry 
hydrophobic core when two or more such chains come together, leading to an 
inter-molecular hydrophobic collapse stabilising the aggregated state.

\begin{acknowledgments}
	We thank the BBSRC, the Frances and Augustus Newman Foundation and the 
	Wellcome Trust (TPJK) for financial support and Daan Frenkel and 
	Suriyanarayanan Vaikuntanathan for helpful discussions.  We are indebted to 
	Michele Vendruscolo for bringing to our attention the data in 
	Fig.~\ref{fig:exp}. DC has been supported in part by Director, Office of 
	Science, Office of Basic Energy Sciences, and by the Division of Chemical 
	Sciences, Geosciences, and Biosciences of the U.S.\ Department of Energy at 
	LBNL under Contract No.\ DE-AC02-05CH11231
\end{acknowledgments}


\begin{thebibliography}{10}

\bibitem{wallqvist1995computer}
Wallqvist, A.; Berne, B. {\em J.~Chem.Phys.} {\bf 1995},
  {\em 99}(9), 2893--2899.

\bibitem{lum1999hydrophobicity}
Lum, K.; Chandler, D.; Weeks, J.~D. {\em J.~Phys.Chem.~B}
  {\bf 1999}, {\em 103}(22), 4570--4577.

\bibitem{huang2001scaling}
Huang, D.~M.; Geissler, P.~L.; Chandler, D. {\em J.~Phys.Chem.~B} {\bf 2001}, 
		{\em 105}(28), 6704--6709.

\bibitem{raschke2001quantification}
Raschke, T.~M.; Tsai, J.; Levitt, M. {\em Proc.~Natl.~Acad.~Sci} {\bf 2001}, 
		{\em 98}(11), 5965--5969.

\bibitem{huang2002hydrophobic}
Huang, D.~M.; Chandler, D. {\em J.~Phys.~Chem.~B} {\bf
  2002}, {\em 106}(8), 2047--2053.

\bibitem{dixit2002molecular}
Dixit, S.; Crain, J.; Poon, W.; Finney, J.; Soper, A. {\em Nature} {\bf 2002},
  {\em 416}(6883), 829--832.

\bibitem{chandler2005interfaces}
Chandler, D. {\em Nature} {\bf 2005}, {\em 437}(7059), 640--647.

\bibitem{granick2008chemistry}
Granick, S.; Bae, S.~C. {\em Science (New York, NY)} {\bf 2008}, {\em
  322}(5907), 1477--1478.

\bibitem{rasaiah2008water}
Rasaiah, J.~C.; Garde, S.; Hummer, G. {\em Annu.~Rev.~Phys.~Chem.} {\bf 2008},
  {\em 59}, 713--740.

\bibitem{WillardChandler2008}
Willard, A.~P.; Chandler, D. {\em J.~Phys.Chem.~B} {\bf
  2008}, {\em 112}(19), 6187--6192.

\bibitem{berne2009dewetting}
Berne, B.~J.; Weeks, J.~D.; Zhou, R. {\em Annu.~Rev.~Phys.~Chem.}
  {\bf 2009}, {\em 60}, 85.

\bibitem{patel2010fluctuations}
Patel, A.~J.; Varilly, P.; Chandler, D. {\em J.~Phys.Chem.~B} {\bf 2010}, {\em 
		114}(4), 1632--1637.

\bibitem{hummer2010molecular}
Hummer, G. {\em Nature chemistry} {\bf 2010}, {\em 2}(11), 906--907.

\bibitem{garde2011unraveling}
Garde, S.; Patel, A.~J. {\em Proc.~Natl.~Acad.~Sci}
  {\bf 2011}, {\em 108}(40), 16491--16492.

\bibitem{patel2012sitting}
Patel, A.~J.; Varilly, P.; Jamadagni, S.~N.; Hagan, M.~F.; Chandler, D.; Garde,
  S. {\em J.~Phys.Chem.~B} {\bf 2012}, {\em 116}(8),
  2498--2503.

\bibitem{ben2015hydrophobic}
Ben-Amotz, D. {\em J.~Phys.~Chem.~Lett} {\bf 2015}.

\bibitem{tenWoldeChandler2002}
ten Wolde, P.~R.; Chandler, D. {\em Proc.~Natl.~Acad.~Sci} {\bf 2002}, {\em 
99}(10), 6539--6543.

\bibitem{miller2007solvent}
Miller, T.~F.; Vanden-Eijnden, E.; Chandler, D. {\em Proc.~Natl.~Acad.~Sci} {\bf 
2007}, {\em 104}(37), 14559--14564.

\bibitem{jamadagni2008interfaces}
Jamadagni, S.~N.; Godawat, R.; Dordick, J.~S.; Garde, S. {\em J.~Phys.Chem.~B} 
{\bf 2008}, {\em 113}(13), 4093--4101.

\bibitem{hummer1996information}
Hummer, G.; Garde, S.; Garcia, A.~E.; Pohorille, A.; Pratt, L.~R. {\em
  Proc.~Natl.~Acad.~Sci} {\bf 1996}, {\em 93}(17),
  8951--8955.

\bibitem{chandler1993gaussian}
Chandler, D. {\em Phys.~Rev.~E} {\bf 1993}, {\em 48}(4), 2898.

\bibitem{varilly2011improved}
Varilly, P.; Patel, A.~J.; Chandler, D. {\em J.~Chem.~Phys.}
  {\bf 2011}, {\em 134}(7), 074109.

\bibitem{Weeks1977}
Weeks, J.~D. {\em J.~Chem.~Phys} {\bf 1977}, {\em 67}, 3106.

\bibitem{Vaikuntanathan2014}
	Vaikuntanathan, S.; Geissler, P.~L. {\em Phys.~Rev.~Lett.}  {\bf 2014}, {\em 
	112}(2), 020603.

\bibitem{Vaikuntanathan2016}
	Vaikuntanathan, S.; Rotskoff, G.; Hudson, A.; Geissler, P.~L. 
	{Proc.~Natl.~Acad.~Sci}  {\bf 2016},
  {\em 113}(16), E2224--E2230.

\bibitem{levitt1976simplified}
Levitt, M. {\em J.~Mol.~Biol.} {\bf 1976}, {\em 104}(1),
  59--107.

\bibitem{Roseman1988}
	Roseman, M.~A. {J.~Mol.~Biol}  {\bf 1988}, {\em 201}(3), 621--623.

\bibitem{Monsellier2007a}
Monsellier, E.; Ramazzotti, M.; de~Laureto, P.~P.; Tartaglia, G.-G.; Taddei,
  N.; Fontana, A.; Vendruscolo, M.; Chiti, F. {\em Biophys.~J.} {\bf
  2007}, {\em 93}(12), 4382 -- 4391.

\bibitem{DuBay2004a}
DuBay, K.~F.; Pawar, A.~P.; Chiti, F.; Zurdo, J.; Dobson, C.~M.; Vendruscolo,
  M. {\em J.~Mol.~Biol}  {\bf 2004}, {\em 341}(5), 1317--1326.

\bibitem{Anfinsen1973}
	Anfinsen, C.~B. {\em Science}  {\bf 1973}, {\em 181}(96), 223--230.

\bibitem{Kyte1982}
Kyte, J.; Doolittle, R.~F. {\em J.~Mol.~Biol} {\bf 1982}, {\em 157}(1), 
105--132.

\end{thebibliography}
\end{document}